\documentstyle[amsfonts,11pt]{article}
\begin{document}
%%%%%%%%%%%%% Front matter %%%%%%%%%%%%%%%%%%%%%%%%%%%%%%%%%%%%%%%%%%%%%%%%
\title{\bf Classical and quasi-classical aspects of supersymmetric
quantum mechanics\thanks{Invited talk presented by G.\ Junker at the
{\it VII International Conference on Symmetry Methods in Physics},
Dubna, July 10--16, 1995.}}
\author{Georg Junker,$^a$
Stephan Matthiesen$^a$ and Akira Inomata$^b$ \\[2mm]
{\it\normalsize $^a$Institut f\"ur Theoretische Physik, Universit\"at
Erlangen-N\"urnberg,}\\ {\it\normalsize Staudtstr.\  7, D-91058 Erlangen,
Germany}\\
{\it\normalsize $^b$Department of Physics, State University of New York,
Albany, N.Y.\ 12222, U.S.A.}}

\date{}
\maketitle

\renewcommand{\baselinestretch}{0.9}
\small
\abstract{\small
A recent development of the studies on classical and quasi-classical
properties of supersymmetric quantum mechanics in Witten's version is
reviewed. First, classical mechanics of a supersymmetric system is
considered. Solutions of the classical equations of motion are given and
their properties are discussed in some detail. The corresponding quantum
model is constructed by canonical quantization. The quantum model is
analyzed by Feynman's path integral within a stationary-phase
approximation. A quasi-classical quantization rule is derived, which is
applicable when supersymmetry is exact or spontaneously broken.}

\renewcommand{\baselinestretch}{1.0}
\normalsize
%%%%%%%%%%%%%%%%%%%%%%%%%%%%%%%%%%%%%%%%%%%%%%%%%%%%%%%%%%%%%%%%%%%%%%%%%%%
%
%%%%%%%%%%%%% Section 1 %%%%%%%%%%%%%%%%%%%%%%%%%%%%%%%%%%%%%%%%%%%%%%%%%%%
\section{Introduction}
The idea of supersymmetry (SUSY) itself is not quite new. It was
originally introduced into physics nearly a quarter century ago
\cite{Mi68,WZ74}. It is based on the assumption that there may
be environments where distinction between bosons and fermions is
irrelevant. A direct and rather successful application of SUSY has been
made in the area of nuclear physics \cite{BBI}. SUSY is essential for the
coherent and self-contained formulation of string theory. SUSY is indeed
a rich idea that has a wide variety of application in physics
\cite{KC85}.

However, as early as 1976, Nicolai \cite{Ni76} suggested that SUSY could
be a tool in non-relativistic quantum mechanics. In 1981, Witten
\cite{Wi81} utilized SUSY quantum mechanics to simulate spontaneous
breaking of SUSY. In the last decade or so, SUSY quantum mechanics has
been extensively studied and widely utilized as a tool for studying
various problems in quantum and statistical physics.

One of the simple non-trivial models of SUSY quantum mechanics is that
of Witten. Besides Witten's original application to SUSY breaking, it
has been used in the study of exactly solvable Schr\"odinger and Dirac
problems, for characterizing certain classical stochastic dynamical
systems via the Fokker-Planck equation, and even in analyzing properties
of semiconductor heterojunctions. For reviews of SUSY methods in quantum
and statistical physics, see, e.g., refs. \cite{SUSYQM,Ju96}.

An interesting property of Witten's model which has attracted
considerable attention is the exactness of modified WKB quantization
formulas \cite{CBC85,IJ91,IJS93} for the so-called shape-invariant
potentials \cite{Gen}. These formulas can be obtained from an evaluation
of Feynman's path integral by the stationary-phase-approximation method.
In contrast to the usual quadratic expansion of the action about the
classical path, the action functional has to be expanded about what we
call the quasi-classical path \cite{IJ91,IJ94}. This brings us to the
question as to what the quasi-classical path is and leads to the study
of quasi-classical mechanics before quantization \cite{JM94}. The aim of
this contribution is to review a recent development of the classical and
quasi-classical properties of SUSY quantum mechanics in Witten's
version.

In Section 2, a classical counterpart of Witten's model is studied.
Solutions of the classical equations of motion are discussed in some
detail. Section 3 describes the quantum version of the classical model
via the canonical quantization. In Section 4, the quasi-classical
quantization formulas are discussed. The pseudoclassical analogue of the
Bohr-Sommerfeld quantization condition is also presented.

%%%%%%%%%%%%%%%%%%%%%%%%%%%%%%%%%%%%%%%%%%%%%%%%%%%%%%%%%%%%%%%%%%%%%%%%%%%
% %%%%%%%%%%%%% Section 2
%%%%%%%%%%%%%%%%%%%%%%%%%%%%%%%%%%%%%%%%%%%%%%%%%%%
\section{Supersymmetric classical mechanics}
Classical supersymmetric models form a subclass of pseudoclassical mechanics,
a notion originally introduced by Casalbuoni \cite{Ca76b}.
Pseudoclassical mechanics deals with classical systems which are
described in terms of Grassmannian variables rather than the usual
Cartesian variables. The degrees of freedom which are described by even
Grassmann numbers correspond to those usually called bosonic degrees of
freedom, whereas those characterized by odd Grassmann numbers give the
fermionic degrees of freedom. These classical models may be viewed as the
classical limits of quantum models with bosonic and fermionic degrees of
freedom \cite{BM75}. In this Section, we shall discuss supersymmetric
classical mechanics, mainly following ref. \cite{JM94}.

Let us start with a pseudoclassical system characterized by the standard
supersymmetric Lagrangian \cite{Ca76c}
\begin{equation}
L:=\frac{1}{2}\dot{x}^2-\frac{1}{2}\Phi ^2(x)+
\frac{{\rm i}}{2}\left(\overline{\psi }\dot{\psi }-
\dot{\overline{\psi }}\psi \right)-\Phi '(x)\overline{\psi }\psi .
\label{L}
\end{equation}
Here $x$ describes a bosonic (i.e.\ an even Grassmann-valued)
degree of freedom, and $\psi $ and $\overline{\psi }$ are fermionic (i.e.\
odd Grassmann-valued) variables. The overbar denotes the Grassmann
variant of complex conjugation. The bosonic and fermionic variables are
elements of the complex Grassmann algebra ${\Bbb C}B_{2}$ generated, for
example, by the two time-independent odd elements $\psi_{0} $,
$\overline{\psi }_{0}$ satisfying the anticommutation relation
$\psi_{0}\overline{\psi}_{0}=-\overline{\psi}_{0}\psi _{0} $:
\begin{equation}
x=a+b\overline{\psi }_{0}\psi _{0} ,\qquad
\psi =c\psi _{0}+d\overline{\psi} _{0} ,\qquad
\overline{\psi} =c^*\overline{\psi} _{0}+d^*\psi _{0},
\label{abcd}
\end{equation}
where $a,b\in{\Bbb R}$ which assures the reality of the bosonic
variable $\overline{x}=x$, and $c, d\in{\Bbb C}$. In (1), $\Phi$ denotes
the so-called SUSY potential which is a differentiable
function of $x$, and $\Phi '(x)={\rm d}\Phi (x)/{\rm d}x$.
As $x$ is an even and real Grassmann number as given in (\ref{abcd}), so is
the potential function. Namely, $\Phi (x)=\Phi (a)+ \Phi '(a)b\overline{\psi
}_{0}\psi _{0}$.

SUSY of the system characterized by (\ref{L}) is obvious. The
following infinitesimal SUSY transformation,
\begin{equation}
\delta x:=\overline{\varepsilon }\psi +\overline{\psi }\varepsilon ,\qquad
\delta \psi:=-({\rm i}\dot{x}+\Phi (x))\varepsilon ,\qquad
\delta \overline{\psi}=({\rm i}\dot{x}-\Phi (x))\overline{\varepsilon},
\label{trafo}
\end{equation}
with infinitesimal odd Grassmann numbers, $\varepsilon $ and
$\overline{\varepsilon }$, gives rise to a variation of the Lagrangian
(\ref{L}):
\begin{equation}
\delta L=\frac{1}{2}\,\frac{{\rm d}}{{\rm d} t}
\left[(\dot{x}-{\rm i}  \Phi )\overline{\varepsilon} \psi +
(\dot{x}+{\rm i} \Phi )\overline{\psi}\varepsilon \right].
\end{equation}
{}From this symmetry follow conserved Noether charges:
\begin{equation}
Q:=\frac{\rm i}{\sqrt{2}}\Bigl(\dot{x}-{\rm i}\Phi(x)\Bigr)\overline{\psi},
\qquad
\overline{Q}=-\frac{\rm i}{\sqrt{2}}\Bigl(\dot{x}+{\rm i}\Phi (x)\Bigr)\psi.
\label{Q}
\end{equation}

The equations of motion which follow from (\ref{L}) are
\begin{equation}
\ddot{x}=-\Phi (x)\Phi '(x)-\Phi ''(x)\overline{\psi }\,\psi ,
\qquad\dot{\psi}=-{\rm i} \Phi '(x)\psi,
\qquad \dot{\overline{\psi }}= {\rm i} \Phi '(x)\overline{\psi} .
\label{eqm}
\end{equation}
{}From the last two equations, it is evident that
$\overline{\psi} (t)\psi (t)=\overline{\psi} (0)\psi (0)$ and
$\psi (t)=\psi (0)\,{\rm e}^{-2{\rm i}\varphi [x]}$
with a phase functional $\varphi [x]$. In fact, the phase coincides with
what we have called the fermionic phase \cite{IJ94},
\begin{equation}
\varphi [x]:={\textstyle\frac{1}{2}}\,\int\limits_{0}^{t}{\rm d}\tau\,
\Phi '\Bigl(x(\tau )\Bigr). \label{pha}
\end{equation}
As in ref. \cite{IJ94}, let us introduce the quasi-classical path,
denoted by $x_{\rm qc}$, as a solution of
\begin{equation}
\ddot{x}=-\Phi (x)\Phi '(x).
\label{xqc}
\end{equation}
If we make an ansatz
$x(t)=x_{\rm qc}(t)+q(t)\overline{\psi }_{0}\psi _{0}$, where $x_{\rm qc}$
and $q$ are real-valued functions of time, and the initial conditions
$\psi (0)=\psi _{0}$, $\overline{\psi} (0)=\overline{\psi} _{0}$ the solutions
of (\ref{eqm}) explicitly read \cite{JM94}
\begin{equation}
\begin{array}{l}
\displaystyle
\psi(t)=\psi_{0}\exp\left\{-2{\rm i}\varphi [x_{\rm qc}]\right\},\qquad
\overline{\psi }(t)=\overline{\psi }_{0}
                    \exp\left\{2{\rm i}\varphi [x_{\rm qc}]\right\}, \\[2mm]
\displaystyle
q(t)=\frac{\dot{x}_{\rm qc}(t)}{\dot{x}_{\rm qc}(0)}\,q(0)
     + \dot{x}_{\rm qc}(t)\int\limits_{0}^{t}{\rm d}\tau
     \frac{F-\Phi '(x_{\rm qc}(\tau ))}{2E-\Phi ^2(x_{\rm qc}(\tau ))}.
\label{sol}
\end{array}
\end{equation}
In the above, $E\geq 0$ and $F\in{\Bbb R}$ are constants of integration.
They are related to the conserved energy ${\cal E}=E+F\overline{\psi
}_{0}\psi _{0}$ associated with the motion of the bosonic degree of
freedom:
\begin{equation}
\begin{array}{rl} {\cal
E}&=\textstyle\frac{1}{2}\dot{x}^2+\frac{1}{2}\Phi ^2(x)+ \Phi
'(x)\overline{\psi }\psi\\[2mm] &=\left[\frac{1}{2}\dot{x}_{\rm
qc}^2+\frac{1}{2}\Phi ^2(x_{\rm qc})\right]+ \Bigl[\dot{x}_{\rm
qc}\dot{q}+\Phi (x_{\rm qc})\Phi '(x_{\rm qc})q+ \Phi '(x_{\rm
qc})\Bigr]\overline{\psi }_{0}\psi _{0}. \end{array}
\end{equation}
Naturally, $E =\frac{1}{2}\dot{x}^2_{{\rm qc}}+
\frac{1}{2}\Phi ^2(x_{\rm qc})$,
is the energy conserved along a quasi-classical path.

The quasi-classical equation of motion (\ref{xqc}) can be derived from either
the Lagrangian of the form,
\begin{equation}
\textstyle
L_{\rm qc}:=\frac{1}{2}\dot{x}^2-\frac{1}{2}\Phi ^2(x)
=\frac{1}{2}\Bigl(\dot{x}\pm{\rm i} \Phi (x)\Bigr)^2\mp{\rm i} \Phi
(x)\dot{x},
\end{equation}
or those of the form,
\begin{equation}
\widetilde{L}^{\pm}_{\rm qc}:=\textstyle\frac{1}{2}\left(\dot{x}\pm{\rm i}
\Phi (x)\right)^2.
\end{equation}
The second set of Lagrangians are interesting in that they are quadratic in
the canonical momentum of $x_{\rm qc}$,
\begin{equation}
\xi ^{\pm}:=\frac{\partial\widetilde{L}^{\pm}_{\rm qc}}{\partial\dot{x}}=
\dot{x}\pm{\rm i} \Phi (x)=(\xi ^\mp)^*,
\end{equation}
which obey along the quasi-classical path the equations $\dot{\xi
}^{\pm}=\mp{\rm i}\Phi '(x_{\rm qc})\xi ^{\pm}$ which are identical in
form to those for the fermionic variables in (\ref{eqm}). Obviously the
solutions can be expressed in terms of the fermionic phase: $\xi
^{\pm}(t)=\xi ^{\pm}(0)\exp\{\mp 2{\rm i}\varphi [x_{\rm qc}]\}$.
However, $\xi ^{\pm}$ are variables complex-valued rather than
Grassmann-valued. We further note that $E=\frac{1}{2}\xi ^+(t)\xi ^-
(t)=const$.  Hence, the fermionic solutions can be put into the form
\begin{equation}
\psi (t)=\frac{1}{2E}\,\xi ^+(0)\xi ^-(t)\psi _{0},\qquad
\overline{\psi}(t)=\frac{1}{2E}\,\xi ^-(0)\xi ^+(t)\overline{\psi}_{0}.
\end{equation}
It is also obvious that the conserved Noether charges (\ref{Q})
are given in the form,
$Q=\frac{{\rm i}}{\sqrt{2}}\,\xi ^{-}(t)\overline{\psi}(t)$,
$\overline{Q}=-\frac{{\rm i}}{\sqrt{2}}\,\xi ^{+}(t)\psi(t)$.
Since the phases of $\xi ^{{-}}(t)$ and $\overline{\psi }(t)$ are equal and
opposite, $Q$ is indeed a constant of motion along the quasi-classical
path. The same can be said for $\overline{Q}$.

Before closing this Section, let us emphasize that the
quasi-classical path $x_{\rm qc}$ completely characterizes the
solutions (\ref{sol}) once initial conditions are given. Another
important point is that along the quasi-classical path $x_{\rm qc}$ the
fermionic phase (\ref{pha}) can be explicitly calculated and put into a
simple form \cite{IJ91,IJ94}:
\begin{equation}
\varphi [x_{\rm qc}]={\textstyle\frac{1}{2}}
\left[\mbox{sgn}(\dot{x}')a(x')-\mbox{sgn}(\dot{x}'')a(x'')\right]+
n_{R}a(x_{R})-n_{L}a(x_{L}).
\label{varphi}
\end{equation}
Here we used the notation $x':=x_{\rm qc}(0)$, $\dot{x}':=\dot{x}_{\rm qc}(0)$,
$x'':=x_{\rm qc}(t)$, $\dot{x}'':=\dot{x}_{\rm qc}(t)$,
$\mbox{sgn}(x):=x/|x|$, $a(x):=\mbox{arcsin}(\Phi (x)/\sqrt{2E})\in [-
\frac{\pi }{2},\frac{\pi }{2}]$, and $n_{R}$ and $n_{L}$ denote the
numbers of right and left turning points along $x_{\rm qc}$, respectively.
These turning points are given by
\begin{equation}
\Phi ^2(x_{R})=2E=\Phi ^2(x_{L}),\qquad x_{R}>x_{L}.
\label{xRL}
\end{equation}

%%%%%%%%%%%%%%%%%%%%%%%%%%%%%%%%%%%%%%%%%%%%%%%%%%%%%%%%%%%%%%%%%%%%%%%%%%%
%
%%%%%%%%%%%%% Section 3 %%%%%%%%%%%%%%%%%%%%%%%%%%%%%%%%%%%%%%%%%%%%%%%%%%%
\section{Supersymmetric quantum mechanics}
In order to quantize the pseudoclassical system (\ref{L}) canonically,
we wish now to move to Hamilton's formulation of mechanics.
The momenta conjugate to the fermionic variables are
\begin{equation}
\Pi :=\frac{\partial L}{\partial\dot{\psi }}=-\frac{\rm i}{2}\,\overline{\psi},
\qquad
\overline{\Pi}:=\frac{\partial L}{\partial\dot{\overline{\psi}}}
=-\frac{\rm i}{2}\,\psi,
\label{Pis}
\end{equation}
which do not depend on $\dot{\psi} $ and $\dot{\overline{\psi }}$.
Hence, the system is subject to the second-class constraints,\footnote{
Here the symbol $\approx$ denotes weak equality following Dirac's
notation \cite{Di64}.}
$\chi _{1}:=\Pi +\frac{\rm i}{2} \,\overline{\psi } \approx 0$ and $\chi
_{2}:=\overline{\Pi} +\frac{\rm i}{2}\,\psi \approx 0$, which have a
non-vanishing Poisson bracket $\{\chi _{1}, \chi _{2}\}_{P} \neq 0$.
The pseudoclassical Hamiltonian, which gives rise to equations of
motion being equivalent to (\ref{eqm}), turns out to be
\begin{equation}
H:=\frac{1}{2}\Bigl(p^{2}+\Phi ^{2}(x)\Bigr)
+{\rm i}\Phi '(x)\left(\overline{\psi }\,\overline{\Pi } -\psi\, \Pi \right).
\label{H}
\end{equation}

Under the constraints, Poisson's brackets are not canonical invariants
in phase space, so that we have to go over to Dirac's bracket formalism
\cite{Di64}. The Dirac brackets for this constrained system have been
calculated \cite{Ma95}, the results being
\begin{equation}
\textstyle
\{x,p\}_{D}=1,
\quad \{\psi ,\Pi \}_{D}=-\frac{1}{2},
\quad \{\overline{\psi} ,\overline{\Pi}\}_{D}=-\frac{1}{2},
\quad \{\psi ,\overline{\psi }\}_{D}=-{\rm i},
\quad \{\Pi ,\overline{\Pi }\}_{D}=\frac{\rm i}{4}
\end{equation}
and all others vanish.

Quantization of this system can be achieved by replacing the $c$-number
variables by the corresponding $q$-number operators and the Dirac
brackets by the corresponding \mbox{(anti-)} commutators divided by
(${\rm i}\hbar$).
Since the fermion variables satisfy the properties, $ \psi^2
=\overline{\psi}^2=
\Pi ^2=\overline{\Pi }^2=0,$ the quantum operators obey the algebra,
\begin{equation}
\begin{array}{l}
[\hat{x},\hat{p}]_{-}={\rm i}\hbar,\qquad
[\hat{\psi },\hat{\Pi }]_{+}=-\frac{{\rm i}\hbar}{2},\qquad
[\hat{\overline{\psi} },\hat{\overline{\Pi} }]_{+}=-\frac{{\rm i}\hbar}{2},
\\[2mm]
[\hat{\psi} ,\hat{\overline{\psi }}]_{+}=\hbar,\qquad
\hat{\psi}^{2}=0=\hat{\overline{\psi}}^{2},\qquad
[\hat{\Pi } ,\hat{\overline{\Pi }}]_{+}=-\frac{\hbar}{4},\qquad
\hat{\Pi}^{2}=0=\hat{\overline{\Pi }}^{2}.
\end{array}
\label{qo}
\end{equation}
Besides the well-known Heisenberg algebra for the bosonic operators
$\hat{x},\hat{p}$, we find, as expected, that $\hat{\psi }$,
$\hat{\overline{\psi }}$, $\hat{\Pi }$ and $\hat{\overline{\Pi }}$
satisfy the algebra of the fermionic creation and annihilation operators.
Obviously, the algebra satisfied by the fermionic operators
is isomophic to that obeyed by Pauli matrices.
Hence, as simple representations of the fermionic operators
we may employ Pauli matrices:
$\hat{\psi }=\sqrt{\hbar}\,\sigma _{-}$,
$\hat{\overline{\psi }}=\sqrt{\hbar}\,\sigma _{+}$,
$\hat{\Pi }=-({\rm i}\sqrt{\hbar}/2)\sigma _{+}$,
$\hat{\overline{\Pi}}=-({\rm i}\sqrt{\hbar}/2)\sigma _{-}$,
and ${\rm i}(\overline{\psi }\,\overline{\Pi }- \psi\,\Pi )=
(\hbar/2)[\sigma _{+},\sigma _{-}]_{-}= (\hbar/2)\sigma _{3}$.
Then we arrive at Witten's quantum mechanical Hamiltonian operator
acting on $L^{2}({\Bbb R})\otimes{\Bbb C}^{2}$:
\begin{equation}
\hat{H}:=\frac{1}{2}\left(\hat{p}^{2}+\Phi ^{2}(\hat{x})+\hbar\Phi '(\hat{x})
\sigma _{3}\right).
\label{hatH}
\end{equation}
Similarly, the conserved charges (\ref{Q}) may be converted into quantum
operators $\hat{Q}$ and $\hat{Q}^{\dagger}$. These charge operators,
together with the Hamiltonian operator (\ref{hatH}), form the
SUSY algebra
\begin{equation}
\left[\hat{Q},\hat{Q}^{\dagger}\right]_{+}=\hat{H},\qquad
\left[\hat{Q},\hat{H}\right]_{-}=0=\left[\hat{Q}^{\dagger},\hat{H}\right]_{-}.
\end{equation}

On the eigenbasis of $\sigma _{3}$ in ${\Bbb C}^{2}$, the Hamiltonian
(\ref{hatH}) becomes diagonal, its diagonal elements being the so-called
partner Hamiltonians acting on $L^{2}({\Bbb R})$:
\begin{equation}
\hat{H}_{\pm}:=
\frac{\hat{p}^{2}}{2}+
\frac{\Phi ^{2}(\hat{x})}{2}\pm\frac{\hbar}{2}\,\Phi '(\hat{x})
\geq 0.
\label{Hpm}
\end{equation}
SUSY of this system explicates itself in the spectral properties of
$\hat{H}_{\pm}$:
\begin{equation}
\mbox{spec}(\hat{H}_{+})/\{0\}=\mbox{spec}(\hat{H}_{-})/\{0\}.
\label{spec}
\end{equation}
Namely, $\hat{H}_{+}$ and $\hat{H}_{-}$ have identical spectra except
for zero.

As in field theory, SUSY is said to be good if there exists an
eigenstate of $\hat{H}$ belonging to a vanishing eigenvalue. Clearly,
this state is an eigenstate of either $\hat{H}_{+}$ or $\hat{H}_{-}$.
SUSY is said broken if such a state does not exist. Whether
SUSY is good or broken depends on the behavior of the SUSY
potential $\Phi $. In order to discriminate between the two cases it is
convenient to utilize the Witten index defined by
\begin{equation}
\Delta :=\dim\mbox{ker}(\hat{H}_{-})-\dim\mbox{ker}(\hat{H}_{+}).
\label{Delta}
\end{equation}
Obviously $\Delta =0$ for broken SUSY, and $\Delta =\pm 1$, if SUSY is
good, depending on whether the state belonging to the zero eigenvalue is
of $\hat{H}_{-}$ or $\hat{H}_{+}$. According to a theorem of Atiyah and
Singer \cite{AS}, the Witten index should be independent of the details of
the SUSY potential $\Phi $. In fact, it depends only on its asymptotic
behavior for $x\to\pm\infty $:
\begin{equation}
\Delta
=\mbox{sgn}\Bigl(\Phi (+\infty )\Bigr)-\mbox{sgn}\Bigl(\Phi (-\infty )\Bigr)
=\frac{1}{\pi }\Bigl(a(x_{R})-a(x_{L})\Bigr).
\label{Delta2}
\end{equation}
For the last equality, we have assumed that there exists a unique set of
left and right turning points for a given energy $E\geq 0$. This is
always the case if $\Phi ^{2}$ has a single-well structure.

%%%%%%%%%%%%%%%%%%%%%%%%%%%%%%%%%%%%%%%%%%%%%%%%%%%%%%%%%%%%%%%%%%%%%%%%%%%
%
%%%%%%%%%%%%% Section 4 %%%%%%%%%%%%%%%%%%%%%%%%%%%%%%%%%%%%%%%%%%%%%%%%%%%
\section{Quasi-classical approximation}
Next, we shall derive quasi-classical formulas for the spectra
of the partner Hamiltonians (\ref{Hpm}).
To this end, we shall utilize Feynman's path integral
for the kernel (the matrix elements of the time-evolution operator)
\cite{IJ91,IJ94}:
\begin{equation}
\langle x''|{\rm e}^{-({\rm i}/\hbar)t\hat{H}_{\pm}}|x'\rangle
=\int{\cal D}x\exp\left\{\frac{\rm i}{\hbar}\,
S_{\rm qc}[x]\mp{\rm i}\varphi [x]\right\},
\label{PI}
\end{equation}
where $S_{\rm qc}[x]:=\frac{1}{2}\int_{0}^{t}{\rm d}\tau (\dot{x}^{2}-
\Phi ^{2}(x))$. Since it is generally difficult to carry out the path
integration of (\ref{PI}), we shall employ the
stationary-phase-approximation method to evaluate it.
A unique feature of our approach is
to expand the action  $S^{\pm}[x]:=S_{\rm qc}[x]\mp\hbar\varphi [x]$ to
second order about the quasi-classical path $x_{\rm qc}$, discussed in
Section 2, rather than the classical path.
To be more explicit, we expand $S_{\rm qc}[x]$
about $x_{\rm qc}$ up to second order in $\eta (\tau ):=x(\tau )-x_{\rm
qc}(\tau )$, so that we have
\begin{equation}
S^{\pm}[x]\simeq S_{\rm qc}[x_{\rm qc}]\mp\hbar\varphi [x_{\rm qc}]+
\frac{1}{2}\int\limits_{0}^{t}{\rm d}\tau
\left[\dot{\eta }^{2}-\left(\Phi ^{2}\right)''
\Bigl(x_{\rm qc}(\tau )\Bigr)\eta ^{2}\right].
\end{equation}
In this approximation, the path integral (\ref{PI}) becomes Gaussian and
can be calculated explicitly \cite{IJ91}. Then we calculate again by the
stationary-phase approximation the kernel of the resolvent
$(E - \hat{H}_{\pm})^{-1}$ given as a Laplace transform of the Feynman
kernel:
\begin{equation}
\langle x''|(E-\hat{H}_{\pm})^{-1}|x'\rangle=\frac{1}{{\rm i}\hbar}
\int\limits_{0}^{\infty }{\rm d}t\,
\langle x''|{\rm e}^{-({\rm i}/\hbar)t\hat{H}_{\pm}}|x'\rangle\,
{\rm e}^{({\rm i}/\hbar)tE},\qquad \mbox{Im}\,E>0.
\end{equation}
As a result, we obtain the supersymmetric version of Gutzwiller's formula
\cite{IJ94}
\begin{equation}
\langle x''|(E-\hat{H}_{\pm})^{-1}|x'\rangle\simeq
\frac{1}{{\rm i}\hbar\,\sqrt{|\dot{x}'\dot{x}''|}}
\sum_{x_{\rm qc}}^{{\rm fixed}\,E}
\exp\left\{\frac{{\rm i}}{\hbar}\,W[x_{\rm qc}]
\mp{\rm i}\varphi [x_{\rm qc}]-{\rm i}(n_{R}+n_{L})\frac{\pi}{2}\right\}
\label{Gutz}
\end{equation}
where $W[x_{\rm qc}]:=\int_{x_{\rm qc}}{\rm d} x\,\sqrt{2E-\Phi
^{2}(x)}$ is Hamilton's characteristic function defined along the
quasi-classical path. The phase $(n_{R}+n_{L})\pi /2$, identified with
the Maslov phase, arises at the turning points of $x_{\rm qc}$. In the
above the sum has to be made over all quasi-classical paths starting
from $x'$ and ending at $x''$ with a fixed energy $E$. Using
the fermionic phase (\ref{varphi}), we may rewrite our
result as follows
\begin{equation}
\begin{array}{rl}
\langle x''|(E-\hat{H}_{\pm})^{-1}|x'\rangle\simeq&\displaystyle
\frac{
\exp\{\mp\frac{\rm i}{2}[\mbox{sgn}(\dot{x}')a(x')-
                         \mbox{sgn}(\dot{x}'')a(x'')]\}}
{{\rm i}\hbar\,\sqrt{|\dot{x}'\dot{x}''|}}\\[2mm]
&\displaystyle\times
\sum_{x_{\rm qc}}^{{\rm fixed}\,E}
\exp\left\{\frac{{\rm i}}{\hbar}\,W[x_{\rm qc}]
-{\rm i}n_{R}\left(\frac{\pi }{2}\pm a(x_{R})\right)
-{\rm i}n_{L}\left(\frac{\pi }{2}\mp a(x_{L})\right)\right\}.
\end{array}
\label{Gutz2}
\end{equation}

{}From the poles of (\ref{Gutz2}) we can obtain quasi-classical
quantization conditions. These poles are easily found by performing the
path sum explicitly. Details are given in \cite{IJ91,IJ94}. Here we only
mention that each complete cycle of a periodic path $x_{\rm qc}$
contributes to the Maslov phase an additive term $\pi $. Similarly, the
contribution of such a cycle to the fermionic phase is given by
\begin{equation}
a(x_{R})-a(x_{L})=\pi \Delta
\end{equation}
as $n_{R}$ and $n_{L}$ increase by one after each cycle (cf.\ relations
(\ref{varphi}) and (\ref{Delta2})). The quasi-classical
supersymmetric quantization condition thus derived reads for $\hat{H}_{\pm}$
\begin{equation}
\oint p_{\rm qc}(x){\rm d}x=
2\pi \hbar \left(n+\frac{1}{2}\pm\frac{\Delta }{2}\right),
\qquad n\in{\Bbb N}_{0},
\label{SWKB}
\end{equation}
where $p_{\rm qc}(x):=\mbox{sgn}(\dot{x})\sqrt{2E-\Phi ^{2}(x)}$ is the
quasi-classical momentum. The integration is to be evaluated for one
cycle of the periodic quasi-classical motion. The additional term $\pi
\hbar $ on the right-hand side stems from the Maslov phase, just as in the
well-known WKB formula, whereas the term $\pi \hbar\Delta$ results from
the fermionic phase. The quantization condition (\ref{SWKB}) coincides
with the formula given in ref.\ \cite{CBC85} if $\Delta =\pm 1$ and the
formula proposed in ref.\ \cite{IJ91,IJ94} if $\Delta =0$. A formula basically
identical to (\ref{SWKB}) was earlier derived by Eckhardt \cite{Ec86}
from a standard WKB consideration. However, he did not recognize the
crucial link between the formula with $\Delta =0$ and broken SUSY.

The quantization condition (\ref{SWKB}), resulting in approximate
energy eigenvalues $E_{n}^{\pm}$, has some remarkable properties.
It leads to the exact ground-state energy, $E_{0}^{-}=0$ for
$\Delta =1$ and $E_{0}^{+}=0$ for $\Delta =-1$, but $E_{0}^{\pm}>0$
for $\Delta =0$. It also gives rise to the relation
$E_{n}^{+}=E_{n+1}^{-}$ for $\Delta =1$, $E_{n}^{-}=E_{n+1}^{+}$ for
$\Delta =-1$ and $E_{n}^{-}=E_{n}^{+}$ for $\Delta =0$. In other words,
the quasi-classical approximation preserves the exact spectral symmetry
(\ref{spec}) between $\hat{H}_{+}$ and $\hat{H}_{-}$. The third and
probably most interesting property, however, is that (\ref{SWKB})
provides exact bound-state spectra for all shape-invariant
potentials (i.e.\ those for which the Schr\"odinger equation
is exactly solvable by the factorization method) \cite{Gen}. In general,
(\ref{SWKB}) provides better approximation. In particular, for broken
SUSY, it has been observed \cite{IJ94} that (\ref{SWKB}) overestimates
the energy eigenvalues, while the usual WKB formula makes
underestimation. Thus, (\ref{SWKB}), together with the standard WKB
approximation, can lead to improved energy spectra.

Finally, we wish to present other aspects of the quasi-classical
quantization condition (\ref{SWKB}). Let us consider the classical phase
integral for the Grassmann variables \cite{Ca76c}
\begin{equation}
\oint\left(\Pi {\rm d}\psi +\overline{\Pi }{\rm d}\overline{\psi }\right)=
\int\limits_{0}^{T_{E}}{\rm d}t\left(\Pi(t)\dot{\psi}(t)+
\overline{\Pi }(t)\dot{\overline{\psi }}(t) \right),
\label{pipsi}
\end{equation}
where $T_{E}:=2\int_{x_{L}}^{x_{R}}{\rm d}x[2E-\Phi ^2(x)]^{-1/2}$ is the
period of the bounded quasi-classical motion. Using the relations (\ref{Pis}),
(\ref{eqm}) and (\ref{Delta2}), we can obtain
\begin{equation}
\begin{array}{ll}
\displaystyle
\oint\left(\Pi {\rm d}\psi +\overline{\Pi }{\rm d}\overline{\psi }\right)&
\displaystyle
=-\frac{1}{2}\int\limits_{0}^{T_{E}}{\rm d}t\,
\Phi '(x_{\rm qc})\left[\overline{\psi }(t),\psi (t)\right]_{-}\\[2mm]
&\displaystyle
=-\Bigl(a(x_{R})-a(x_{L})\Bigr)\left[\overline{\psi }_{0},\psi _{0}\right]_{-}
=-\pi \Delta \left[\overline{\psi }_{0},\psi _{0}\right]_{-}.
\end{array}
\label{new}
\end{equation}
This is a very interesting result. It shows that the Witten index
(\ref{Delta}) can be directly related to
the phase integral (\ref{pipsi}) which is purely classical. It
has been shown by Ma\~{n}es and Zumino \cite{MZ86} that the Witten index
can be derived from a pseudo-classical generalization of Van Vleck's
formula. These authors have noted that for an evaluation of the index only
classical quantities are need. Nevertheless, their formula (67) in
\cite{MZ86} is still a quantum mechanical expression containing $\hbar$.
In contrast, the present formula (\ref{new}) expresses the Witten index
in terms of purely classical quantities.

It is also entertaining to incorporate (\ref{new}) into (\ref{SWKB}).
Following the canonical quantization procedure shown in Section 3, let
us replace the commutator of the classical Grassmann numbers in
(\ref{new}) by $\hbar\sigma _{3}$. Since $\sigma _{3}$ has eigenvalues
$\pm 1$ in the subspaces corresponding to $\hat{H}^{\pm}$,
the quasi-classical quantization formula (\ref{SWKB}) may formally be put into
the form
\begin{equation}
\oint\left(p_{\rm qc}{\rm d}x+\Pi {\rm d}\psi +
\overline{\Pi }{\rm d}\overline{\psi }\right)
=2\pi \hbar\left(n+\frac{1}{2}\right).
\end{equation}
This formula is the pseudoclassical analogue of the Bohr-Sommerfeld
quantization condition.

%%%%%%%%%%%%%%%%%%%%%%%%%%%%%%%%%%%%%%%%%%%%%%%%%%%%%%%%%%%%%%%%%%%%%%%%%%%
%
%%%%%%%%%%%%% Acknowledgements %%%%%%%%%%%%%%%%%%%%%%%%%%%%%%%%%%%%%%%%%%%%
\section*{Acknowledgements}
Two of us (AI and GJ) would like to thank the organizers for their kind
invitation to this conference. One of us (GJ) has been supported by
the Heisenberg-Landau program and the Deutsche Forschungsgemeinschaft, which
is gratefully acknowledged.

%%%%%%%%%%%%%%%%%%%%%%%%%%%%%%%%%%%%%%%%%%%%%%%%%%%%%%%%%%%%%%%%%%%%%%%%%%%
%
%%%%%%%%%%%%% References %%%%%%%%%%%%%%%%%%%%%%%%%%%%%%%%%%%%%%%%%%%%%%%%%%

%%%%%%%%%%%%%%%%%%%%%%%%%%%%%%%%%%%%%%%%%%%%%%%%%%%%%%%%%%%%%%%%%%%%%%%%%%%

\begin{thebibliography}{99}
\itemsep 0mm
\small
\bibitem{Mi68}H.\ Miyazawa, Phys.\ Rev.\ {\bf 170} (1968) 1596;
Y.A.\ Gol'fand and E.P.\ Likhtam, Pis'ma Zh.\ Eksp.\ Teor.\ Fiz.\ {\bf 13}
(1971) 452 [JETP Lett.\ {\bf 13} (1971) 323];
D.V.\ Volkov and V.P.\ Akulov, {\it ibid}.\ {\bf 16} (1972) 621
[{\bf 16} (1972) 438].
\bibitem{WZ74}J.\ Wess and B.\ Zumino, Nucl.\ Phys.\ {\bf B70} (1974)
39, {\it ibid}.\ {\bf B78} (1974) 1.
\bibitem{BBI}A.B.\ Balantekin, I.\ Bars and F.\ Iachello, Phys.\ Rev.\
Lett.\ {\bf 47} (1981) 19.
\bibitem{KC85}V.A.\ Kosteleck\'{y} and  D.K.\ Campbell eds., {\it
Supersymmetry in Physics}, (North-Holland, Amster\-dam, 1985).
\bibitem{Ni76}H.\ Nicolai, J.\ Phys.\ A {\bf 9} (1976) 1497.
\bibitem{Wi81}E.\ Witten, Nucl.\ Phys.\ {\bf B188} (1981) 513.
\bibitem{SUSYQM}A.\ Lahiri, P.K.\ Roy and B.\ Bagchi, Int.\ J.\ Mod.\
Phys.\ A {\bf 5} (1990) 1383; B.\ Roy, P.\ Roy and R.\ Roychoudhury,
Fortschr.\ Phys.\ {\bf 89} (1991) 211; F.\ Cooper, A.\ Khare and U.\
Sukhatme, Phys.\ Rep.\ {\bf 251} (1995) 267.
\bibitem{Ju96}G.\ Junker, {\it Supersymmetric methods in quantum and
statistical physics}, (Springer-Verlag, Berlin, in preparation).
\bibitem{CBC85}A.\ Comtet, A.D.\ Bandrauk and D.K.\ Campbell, Phys.\ Lett.\ B
{\bf 150} (1985) 159.
\bibitem{IJ91}A.\ Inomata and G.\ Junker, in H.\
Cerdeira, S.\ Lundqvist, D.\ Mugnai, A.\ Ranfagni, V.\ Sa-yakanit and L.S.\
Schulman eds., {\it Lectures on Path Integration: Trieste 1991}, (World
Scientific, Singapore, 1993) p.460; also in J.Q.\ Liang, M.L.\ Wang,
S.N.\ Qiao and D.C.\ Su eds., {\it Proceedings of International Symposium
on Advanced Topics of Quantum Physics}, (Science Press, Beijing, 1993)
p.61.
\bibitem{IJS93}A.\ Inomata, G.\ Junker and A.\ Suparmi, J.\ Phys.\ A
{\bf 26} (1993) 2261.
\bibitem{Gen}L.\'E.\ Gendenshtein, Pis'ma Zh.\ Eksp.\ Teor.\ Fiz.\ {\bf 38}
(1983) 299 [JETP Lett.\ {\bf 38} (1983) 356].
\bibitem{IJ94}A.\ Inomata and G.\ Junker, Phys.\ Rev.\ A {\bf 50} (1994)
3638.
\bibitem{JM94}G.\ Junker and S.\ Matthiesen, J.\ Phys.\ A {\bf 27} (1994)
L751, Addendum {\it ibid}.\ {\bf 28} (1995) 1467.
\bibitem{Ca76b}R.\ Casalbuoni, Nuovo Cim.\ {\bf 33A} (1976) 115.
\bibitem{BM75} F.A.\ Berezin and M.S.\ Marinov, Pis'ma Zh.\ Eksp.\ Teor.\
Fiz.\ {\bf 21} (1975) 678 [JETP Lett.\ {\bf 21} (1975) 320];
Ann.\ Phys.\ (NY) {\bf 104} (1977) 336.
\bibitem{Ca76c}R.\ Casalbuoni, Nuovo Cim.\ {\bf 33A} (1976) 389.
\bibitem{Ec86}B.\ Eckhardt, Phys.\ Lett.\ {\bf 168B} (1986) 245.
\bibitem{Di64}P.A.M.\ Dirac, {\it Lectures on Quantum Mechanics},
(Yeshiva Univ., New York, 1964).
\bibitem{Ma95}S.\ Matthiesen, diploma thesis,
University of Erlangen-N\"urnberg, 1995.
\bibitem{AS}M.F.\ Atiyah and I.M.\ Singer, Bull.\ Amer.\ Math.\ Soc.\ {\bf 69}
(1963) 422.
\bibitem{MZ86}J.\ Ma\~{n}es and B.\ Zumino, Nucl.\ Phys.\ {\bf B270} [FS16]
(1986) 651.
\end{thebibliography}
\end{document}